\documentclass[conference]{IEEEtran}
\IEEEoverridecommandlockouts
\usepackage{cite}
\usepackage{amsmath,amssymb,amsfonts}
\usepackage{algorithmic}
\usepackage{graphicx}
\usepackage{textcomp}
\usepackage{tabularx}
\usepackage{xcolor}
\def\BibTeX{{\rm B\kern-.05em{\sc i\kern-.025em b}\kern-.08em
    T\kern-.1667em\lower.7ex\hbox{E}\kern-.125emX}}
\begin{document}

\title{Leveraging Peer, Self, and Teacher Assessments for Generative AI-Enhanced Feedback\\
}

\author{\IEEEauthorblockN{Alvaro Becerra}
\IEEEauthorblockA{\textit{Department of Computer Science Engineering} \\
\textit{Universidad Autónoma de Madrid}\\
Madrid, Spain \\
alvaro.becerra@uam.es}
\and
\IEEEauthorblockN{Ruth Cobos}
\IEEEauthorblockA{\textit{Department of Computer Science Engineering} \\
\textit{Universidad Autónoma de Madrid}\\
Madrid, Spain \\
ruth.cobos@uam.es}
}

\maketitle

\begin{abstract}
Providing timely and meaningful feedback remains a persistent challenge in higher education, especially in large courses where teachers must balance formative depth with scalability. Recent advances in Generative Artificial Intelligence (GenAI) offer new opportunities to support feedback processes while maintaining human oversight. This paper presents an study conducted within the AICoFe (AI-based Collaborative Feedback) system, which integrates teacher, peer, and self-assessments of engineering students’ oral presentations. Using a validated rubric, 46 evaluation sets were analyzed to examine agreement, correlation, and bias across evaluators. The analyses revealed consistent overall alignment among sources but also systematic variations in scoring behavior, reflecting distinct evaluative perspectives. These findings informed the proposal of an enhanced GenAI model within AICoFe system, designed to integrate human assessments through weighted input aggregation, bias detection, and context-aware feedback generation. The study contributes empirical evidence and design principles for developing GenAI-based feedback systems that combine data-based efficiency with pedagogical validity and transparency.

\end{abstract}


\section{Introduction}
Formative assessment and feedback are widely recognized as fundamental components of effective learning in higher education. Rather than serving merely as a mechanism for evaluation, feedback functions as a dynamic process that fosters learning, reflection, and performance improvement \cite{morris2021formative}. However, this process cannot rely solely on teacher-provided judgments; it requires authentic evaluative experiences in which students actively engage in giving and receiving feedback \cite{sadler1989formative}.

A valuable way to engage students in this evaluative process is through self-assessment, which involves reflecting on and judging the quality of one’s own work against explicit criteria. In particular, when students practice criterion-referenced self-assessment, they become more aware of learning goals, better understand teachers’ expectations, and use the process to identify strengths and areas for improvement. Beyond enhancing performance, self-assessment promotes metacognition and self-regulated learning, as students internalize standards of quality and develop a sense of responsibility for their progress \cite{andrade2007student}.

Complementary to this, peer assessment invites students to provide constructive feedback on the work of their classmates. This approach not only helps students recognize errors and refine their work but also allows evaluators to apply shared criteria critically, deepening their understanding and avoiding similar mistakes in their own tasks \cite{topping2009peer}. However, peer feedback also presents several well-documented challenges. Students providing feedback often struggle to compose comments that are sufficiently detailed, balanced, and constructive, partly because of limited experience, uncertainty about quality standards, or a lack of confidence in their evaluative abilities. At the same time, feedback recipients may find it difficult to interpret and apply peer comments, particularly when these address superficial features rather than the substance of the work \cite{wei2024incorporating}.

To address these challenges, the integration of Generative Artificial Intelligence (GenAI) offers promising opportunities to enhance feedback processes. For instance, \cite{sajadi2024harnessing} has shown that incorporating tools such as ChatGPT into peer review activities can make feedback more actionable, specific, and constructive, while simultaneously filtering out unhelpful or inappropriate comments. In this context, the AICoFe system (``Artificial Intelligence-based Collaborative Feedback System'') \cite{becerra2025enhancing}, developed at Universidad Autonoma de Madrid, integrates rubric-based peer and teacher assessments with a Generative AI model (GePeTo \cite{becerra2024generative}) to generate personalized, high-quality feedback. In addition, AICoFe enables students to perform self-assessment, fostering reflection and the development of both technical (hard) and professional (soft) skills.

However, a current limitation of AICoFe and similar AI-enhanced feedback systems lies in their equal weighting of evaluative sources, typically treating teacher and peer inputs as equally reliable, without considering potential variations in bias, consistency, or perspective among them. Moreover, self-assessment data are often underutilized or entirely excluded, despite their recognized value in promoting self-regulated learning and evaluative judgment.

Building on these considerations, this study contributes to advancing the design of Generative AI–enhanced feedback systems through the following key contributions:
\begin{itemize}
    \item \textbf{Comprehensive analysis of human evaluative sources:} We examine the consistency, agreement, and interrelations among teacher, peer, and self-assessments within the AICoFe system, using a validated rubric for oral presentations in engineering education.
    \item \textbf{Empirical evidence of biases and alignment patterns:} We identify systematic differences in scoring behavior across evaluators and analyze their relationship with the teacher’s overall grade to assess the reliability and validity of each feedback source.
    \item \textbf{Pedagogical and design implications for GenAI feedback:} Based on the results, we discuss directions to enhance the GePeTo generative model within AICoFe system by incorporating weighted feedback integration and self-assessment.
\end{itemize}

The remainder of this paper is organized as follows. Section \ref{s:related} reviews prior research on rubrics, peer, self-, and teacher assessments, and the integration of GenAI in educational feedback systems. Section \ref{s:metho} describes the methodology of our study, including the context, participants, assessment instruments, data collection procedures, and statistical analyses. Section \ref{s:results} presents the results. Section \ref{s:discussion} discusses the findings in relation to existing literature and highlights design implications for AI-enhanced feedback within the AICoFe system. Finally, Section \ref{s:conclusions} concludes the paper and outlines directions for future research.

\section{Related Work}\label{s:related}

\subsection{Rubrics in Peer, Self, and Teacher Assessment}

Teacher assessment has traditionally been at the core of educational evaluation, as teachers are primarily responsible for designing, implementing, and interpreting assessment practices that guide student learning \cite{black2018classroom, brown2022past}. Beyond grading, teacher assessment assumes a formative dimension by informing instructional decisions and providing feedback that promotes student reflection and improvement. Within this context, several studies have highlighted the value of rubrics as essential tools for supporting teachers in articulating learning goals, communicating quality criteria, and ensuring consistency and transparency in feedback \cite{pastore2023teacher}.

When effectively designed, rubrics can serve as powerful formative instruments that guide teachers’ qualitative judgments rather than simply quantifying performance. This qualitative orientation enables teachers to provide feedback that is descriptive, constructive, and improvement-oriented instead of relying on aggregated scores. Moreover, the formative use of rubrics helps clarify learning expectations, reduce ambiguity in evaluation, and foster meaningful dialogue between teachers and students about quality standards and progress \cite{panadero2020critical, brookhart2018appropriate}.

By making learning criteria explicit and fostering a shared understanding of quality, rubrics not only enhance the consistency of teacher assessment but also lay the foundation for more participatory forms of evaluation. When students are familiar with the criteria used by their teachers, they are better equipped to apply these same standards to assess their own work and that of their peers. In this sense, rubrics act as mediating tools that bridge teacher, self-, and peer assessment, promoting a common evaluative language and supporting the development of students’ evaluative judgment and self-regulated learning \cite{panadero2020critical, andrade2007student}. Likewise, rubrics provide clear reference points that help students interpret quality, offer more focused feedback, and reflect critically on their own performance \cite{reinholz2016assessment}.

\subsection{Comparing Peer, Self, and Teacher Assessment}
The relationship among peer, self-, and teacher assessment has been widely examined in higher education to explore issues of validity, reliability, and bias. Early meta-analyses established that peer ratings tend to approximate teacher ratings to a moderate degree. For instance, \cite{falchikov2000student} reported an average correlation of $r = 0.69$ across 48 studies in higher education. Their findings highlighted that peer marks align more closely with teacher marks when global judgments based on clearly defined criteria are used, as opposed to fragmented, dimension-specific ratings. Similarly, \cite{li2016peer} reported slightly lower but still substantial average correlation of $r = 0.63$ between peer and teacher ratings, confirming that peer assessment can produce reasonably valid results, even in online settings. However, their analysis revealed important moderating factors: peer–teacher agreement is stronger in graduate courses, for individual rather than group work, when assessors are non-anonymous, when feedback includes qualitative comments, and when students are involved in developing rating criteria.

Despite these generally positive findings, several studies point to systematic biases in peer evaluation, such as leniency, friendship, or reciprocity effects, which can undermine reliability \cite{magin2001reciprocity, stonewall2018review}. While \cite{magin2001reciprocity} demonstrated that reciprocity effects account for only a small portion of score variance, later studies highlight the persistence of social and demographic biases, such as gender and cultural stereotypes, particularly in collaborative and team-based learning environments \cite{stonewall2018review}.

Comparatively, self-assessment research shows a more nuanced pattern. While self-ratings often diverge from teacher marks, especially among less experienced students, they strongly contribute to metacognitive development and self-regulated learning \cite{ross2006reliability, nieminen2025student}. Recent meta-reviews confirm that self-assessment accuracy improves when rubrics are used and when students receive guided reflection opportunities \cite{tang2023comparison}.

\subsection{GenAI-Supported Feedback and Learning Analytics}

Recent research in educational technology have emphasized the transformative potential of Generative Artificial Intelligence (GenAI) in supporting feedback and Learning Analytics. For instance, \cite{giannakos2025promise} highlights that large language models (LLMs) can be leveraged to diagnose learners’ challenges, generate personalized feedback, and enhance self-regulated learning processes, although the authors also warn of significant limitations regarding reliability, ethics, and human oversight.

This growing interest in GenAI-supported learning has led to the rapid development of various educational tools that integrate multimodal generative models. Recent studies indicate that such systems can process multiple input types (e.g., text, image, audio, and video) to provide adaptive feedback, automated assessment, and personalized learning experiences \cite{imran2024google}. For example, \cite{mello2025empowering} presented an approach that leverages Large Language Models (LLMs) to transcribe handwritten student essays and automatically score them according to predefined educational criteria. Their study demonstrates how multimodal LLMs can convert handwritten text into digital form and evaluate writing quality, providing feedback while reducing teachers’ workload. Similarly, \cite{xavier2025empowering} evaluated an AI-driven feedback platform designed to assist instructors in assessing open-ended student responses. The system integrates LLMs and learning analytics to generate personalized, tag-based feedback and recommendations, improving both the quality and efficiency of the assessment process. Their findings show that instructors not only produced more detailed feedback with LLM support but also adopted most of the AI-generated suggestions with minimal modification.

Learning Analytics dashboards \cite{verbert2020learning,becerra2023m2lads} have also evolved through the incorporation of generative AI technologies. For example, \cite{yan2024vizchat} introduced VizChat, a prototype designed to augment traditional dashboards with contextualised explanations generated by multimodal generative AI. VizChat enables users to interact with dashboard visualisations through natural language, receiving personalised and context-aware explanations of complex learning data. This approach addresses the cognitive overload often associated with visual analytics by allowing learners and instructors to query visualisations conversationally, enhancing comprehension and reflective practice. Similarly, \cite{becerra2024generative} proposed GePeTo, a generative AI-based personalized guidance tool integrated into dashboards to provide students with adaptive feedback and support. GePeTo leverages anonymized data and indicators to generate customized guidance messages validated for accuracy and tone before being displayed in the dashboards.

However, as \cite{topali2025designing} point out, the growing use of GenAI and AI-based systems in education requires a stronger human-centered and pedagogically grounded perspective to ensure that these technologies align with pedagogical values and ethical principles.

\section{Methodology}\label{s:metho}

\subsection{Context and Participants} 
The study \cite{becerra2026multimodal} involved 46 undergraduate students (6 female students) in the final year of the Telecommunication Technology and Service Engineering program at Universidad Autonoma de Madrid. They were enrolled in a compulsory course to train students in enhancing communication and oral presentation skills in engineering contexts. 

Each student delivered one individual presentation of approximately 10 minutes, followed by a 5-minute Q\&A session. The presentation topics were related to engineering and technology, and were randomly assigned from a predefined list. 

\begin{figure}[h]
    \centering
    \includegraphics[width=\linewidth]{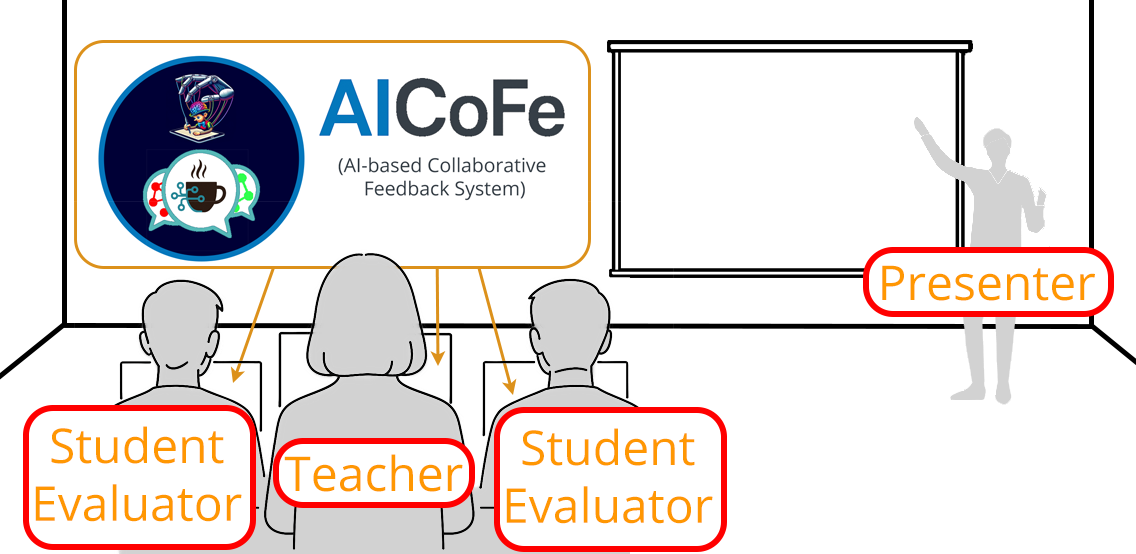}
\caption{Classroom setup of the peer assessment activity using the AICoFe system}
    \label{fig:setup}
\end{figure}

During the course, a peer assessment methodology was implemented using the AICoFe system \cite{becerra2025enhancing}. As illustrated in Figure \ref{fig:setup}, students had two roles: evaluator (two peers per presentation) and presenter. The course teacher also acted as evaluator, using the same rubric implemented in AICoFe. All presentations were delivered in a face-to-face classroom setting. For each presentation, the two peer evaluators were randomly assigned to ensure fairness and to avoid systematic bias. 

After each presentation, students performed a self-assessment using the same rubric in AICoFe while reviewing a video recording of their own performance.

\subsection{Assessment Instrument (Rubric)} 
The presentations were assessed using a rubric implemented in the AICoFe system. The rubric comprised 15 items divided into three main sections: (1) Content, (2) Design, and (3) Delivery. Each item was scored on a five-point Likert scale (1 = very poor, 5 = excellent). To ensure consistency in the evaluations, each level of the scale included a brief description of its meaning. Examples of rubric items and their scale levels are provided in Table~\ref{tab:rubric_examples}.

The rubric was validated by the course teacher and external experts in oral communication. It was introduced and explained to students at the beginning of the course, and both students and the teacher were able to access it at any time within the AICoFe system. In addition to quantitative scoring, the rubric also allowed for qualitative comments, as evaluators could provide an open-ended observation for each item.

\begin{table*}[h]
\centering
\caption{Examples of rubric items and performance levels}
\label{tab:rubric_examples}
\renewcommand{\arraystretch}{1.3} 
\begin{tabularx}{\textwidth}{|l|X|X|X|X|X|}
\hline
\textbf{Item} & \textbf{5 – Excellent} & \textbf{4 – Good} & \textbf{3 – Fair} & \textbf{2 – Poor} & \textbf{1 – Very Poor} \\ \hline

\textbf{Opening of the presentation} & 
Clear and impactful introduction, captures attention and defines the topic well. & 
Adequate start, provides context but lacks impact. & 
Relevant information but fails to engage the audience. & 
Confusing start or weak connection with the topic. & 
No clear introduction, lacks structure, purpose not defined. \\ \hline

\textbf{Text readability} & 
Appropriate font size and style, consistent throughout slides. & 
Correct size, minor inconsistencies in font or format. & 
Legible font, but with errors in size or formatting. & 
Inadequate size or notable inconsistencies. & 
Illegible text or very poor formatting. \\ \hline

\textbf{Body language} & 
Natural and expressive, conveys confidence. & 
Good posture, occasional lack of gestures. & 
Correct posture but limited gestures. & 
Unnatural movements or closed posture. & 
Inadequate body language, conveys insecurity. \\ \hline

\end{tabularx}
\end{table*}

\subsection{Data Collection Procedure} 
All evaluation data were collected through the AICoFe system. During each session, presentations were evaluated in real time by the teacher and two randomly assigned peers using the online rubric within AICoFe. Then, students completed a self-assessment while reviewing a video recording of their own presentation. This sequential process ensured that all roles (teacher, peer, and self) used the same instrument under comparable conditions. The AICoFe system stored all evaluations and allowed exporting the dataset for further statistical analysis. In addition to the rubric-based evaluation, the teacher also assigned a global grade for each presentation, which was determined independently and not derived from the rubric. This global grade was reported on a 0–10 scale, consistent with the institutional grading system. 

A total of 46 full evaluation sets were collected, each including the teacher’s rubric-based scores, the teacher’s overall grade (assigned independently and not derived from the rubric), two peer evaluations, and one self-evaluation. 

\subsection{Statistical Analyses} 
Statistical analyses were conducted to examine the consistency and agreement between different evaluators (teacher, peers, and self-assessment), as well as their relationship with the teacher’s overall grade.

To measure inter-rater agreement, we employed several complementary indices. Weighted Cohen’s kappa was used to estimate pairwise agreement, although its known sensitivity to skewed distributions motivated the additional use of Gwet’s AC2 coefficient, which is more robust in such cases \cite{vach2023gwet}. Kendall’s coefficient of concordance (W) was calculated to assess the global agreement across all evaluators. 

The relationship between rubric scores and the teacher’s overall grade was examined using correlation analyses. Since the analyses were based on average rubric scores (e.g., mean of all items for each evaluator or group of evaluators), these values can be considered continuous variables, which justifies the use of Pearson’s correlation coefficient. Spearman’s rank correlation was also computed as a complementary measure.

Potential systematic biases in scoring were explored through paired-sample t-tests, comparing mean scores across evaluator groups. Normality of score differences was assessed using Shapiro–Wilk tests, and either paired $t$-tests or Wilcoxon signed-rank tests were applied accordingly.

All statistical tests were performed with a significance level of $\alpha = 0.05$. The analyses were carried out in Python, using \texttt{pandas}, \texttt{scipy}, \texttt{sklearn}, \texttt{pingouin} and \texttt{irrCAC}.

\subsection{Ethical Considerations} 
Participation in the peer assessment activity was a mandatory component of the course, as it formed part of the teaching methodology. However, students were informed that the use of the collected data for research purposes was voluntary, and their course grades were not affected by their decision to allow the data to be included in the study. Importantly, the grades that students received in the course were not influenced by the scores assigned by their peers in the rubric; only the teacher’s overall grade determined the final course outcome. Peer evaluations were anonymized within the AICoFe system to reduce potential bias, and all data were handled in compliance with institutional ethical guidelines.  

The entire experimental setup was conducted within the framework of a teaching innovation project officially approved and supported by the university, which ensured adherence to both pedagogical and ethical standards.

\section{Results}\label{s:results}
\subsection{Agreement Across Evaluators} 
Table~\ref{tab:agreement} summarizes the average inter-rater agreement indices across all presentations. Weighted Cohen’s $\kappa$ values indicated weak levels of agreement, ranging from 0.22 (teacher vs. self-assessment) to 0.30 (peers agreement). In contrast, Gwet’s AC2 consistently yielded higher values, between 0.58 and 0.63, suggesting moderate to substantial agreement. Exact agreement percentages varied between 0.43 and 0.53 across evaluator pairs. 

The overall concordance among evaluators was moderate. Kendall’s coefficient of concordance (W) reached 0.51 for the teacher and peers, and 0.46 when including self-assessments. Global AC2 values were similar, with 0.66 in both cases (Table~\ref{tab:kendall}). These results indicate that, although perfect consistency was not achieved, there was a meaningful level of agreement between evaluators.

\begin{table}[h]
\centering
\caption{Mean pairwise agreement between evaluators}
\label{tab:agreement}
\begin{tabularx}{\columnwidth}{Xccc}
\hline
\textbf{Comparison} & \textbf{$\kappa_w$} & \textbf{AC2} & \textbf{Exact agreement} \\
\hline
Peer 1 vs Peer 2                 & 0.30 & 0.63 & 0.53 \\
Teacher vs mean of Peers         & 0.23 & 0.60 & 0.46 \\
Teacher vs Self-assessment       & 0.22 & 0.58 & 0.43 \\
Mean of Peers vs Self-assessment & 0.27 & 0.63 & 0.47 \\
\hline
\end{tabularx}
\end{table}

\begin{table}[h]
\centering
\caption{Overall agreement among all evaluators}
\label{tab:kendall}
\begin{tabular}{lcc}
\hline
\textbf{Evaluator group} & \textbf{Kendall’s W} & \textbf{Gwet’s AC2} \\
\hline
Teacher and Peers             & 0.51 & 0.66 \\
Teacher, Peers, and Self-assessment      & 0.46 & 0.66 \\
\hline
\end{tabular}
\end{table}

\subsection{Correlation Across Evaluations} 

\begin{figure*}[ht]
    \centering
    \includegraphics[width=\linewidth]{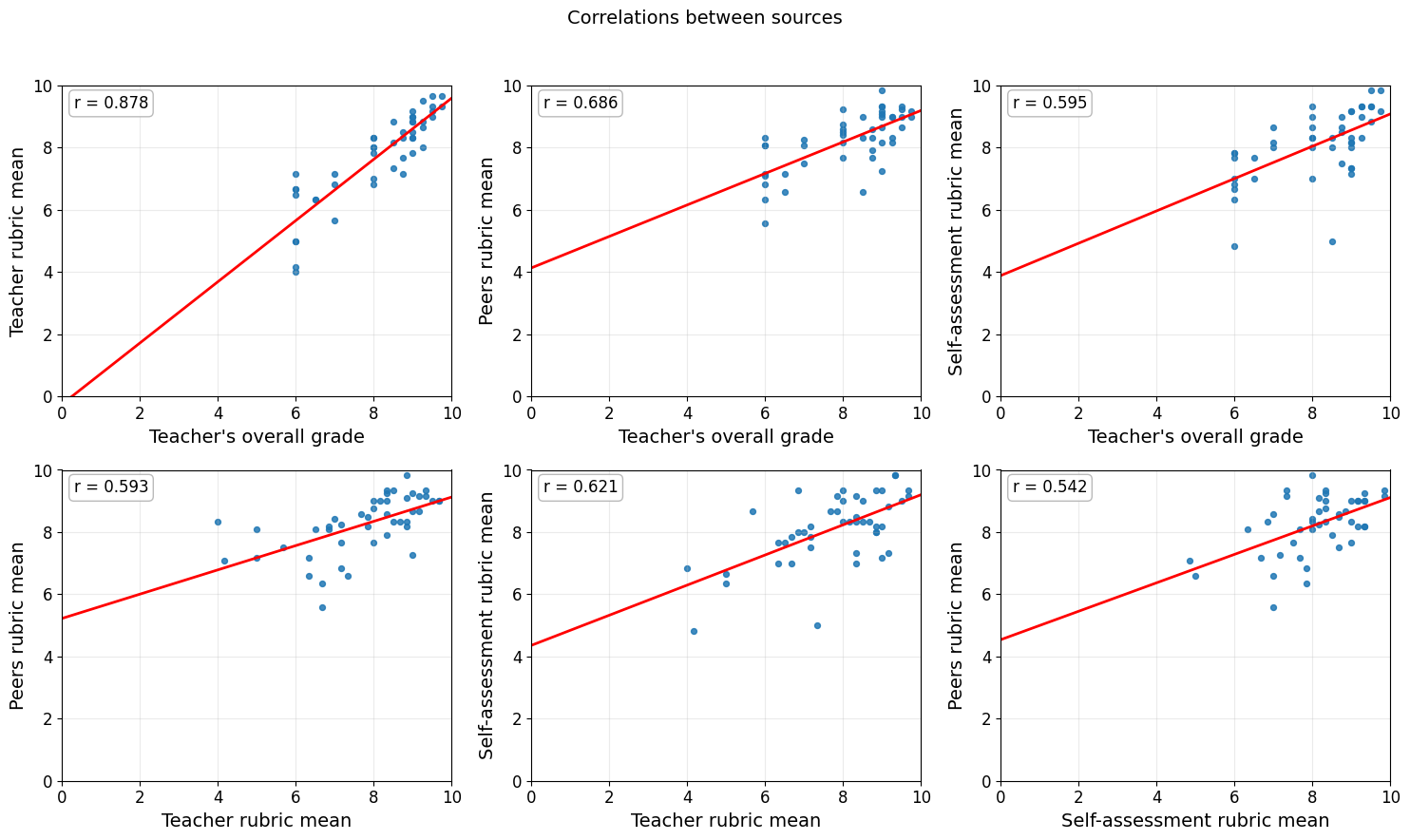}
    \caption{Pairwise correlations between rubric-based and teacher's overall grades across evaluators (teacher, peers, and self-assessments). Each subplot displays the Pearson correlation coefficient ($r$) and the regression line (in red)}
    \label{fig:correlations}
\end{figure*}
Correlations between rubric-based scores and the teacher’s overall grade are presented in Table~\ref{tab:correlation}. Both Pearson’s $r$ and Spearman’s $\rho$ coefficients were computed. The strongest associations were found between the teacher’s own rubric scores and her overall grade ($r = 0.88$, $\rho = 0.89$). The mean rubric score across all evaluators (teacher, peers, and self-assessment) also correlated strongly with the teacher's overall grade ($r = 0.85$, $\rho = 0.86$). 

Peer assessments showed a lower, though still significant, correlation with the teacher’s overall grade ($r = 0.69$, $\rho = 0.68$). Self-assessments were the least aligned with the teacher’s overall grading, with moderate correlations ($r = 0.59$, $\rho = 0.67$). All correlations were statistically significant ($p < 0.001$).

\begin{table}[h]
\centering
\caption{Correlations between rubric-based scores and teacher’s overall grade ($n=46$)}
\label{tab:correlation}
\begin{tabular}{lcc}
\hline
\textbf{Score type} & \textbf{Pearson’s $r$} & \textbf{Spearman’s $\rho$} \\
\hline
Teacher rubric mean                  & 0.88 & 0.89 \\
Peers rubric mean                    & 0.69 & 0.68 \\
Mean rubric score (Teacher + Peers)  & 0.86 & 0.83 \\
Self-assessment rubric mean          & 0.59 & 0.67 \\
Mean rubric score (All evaluators)   & 0.85 & 0.86 \\
\hline
\end{tabular}
\end{table}

Table~\ref{tab:correlation_rubrics} presents the correlations among rubric-based mean scores across evaluator types. All coefficients were positive and statistically significant ($p < 0.001$). The strongest association was found between the teacher’s and self-assessment rubric means ($r = 0.62$), followed by the teacher–peers ($r = 0.59$) and peers–self ($r = 0.54$) comparisons.

Figure~\ref{fig:correlations} displays the pairwise correlations between rubric-based and teacher's overall grades across evaluators.

\begin{table}[h]
\centering
\caption{Correlations between rubric-based scores across evaluator types ($n=46$)}
\label{tab:correlation_rubrics}
\begin{tabularx}{\columnwidth}{Xcc}
\hline
\textbf{Comparison} & \textbf{Pearson’s $r$} & \textbf{Spearman’s $\rho$} \\
\hline
Teacher rubric mean vs. Peers rubric mean        & 0.59 & 0.69 \\
Teacher rubric mean vs. Self-assessment rubric mean & 0.62 & 0.56 \\
Peers rubric mean vs. Self-assessment rubric mean   & 0.54 & 0.49 \\
\hline
\end{tabularx}
\end{table}

\subsection{Bias Analysis} 
Paired-sample tests were conducted to examine potential systematic differences between evaluators’ rubric scores and the teacher’s overall grade. Results are summarized in Table~\ref{tab:bias}. 

\begin{figure}[th]
    \centering
    \includegraphics[width=\linewidth]{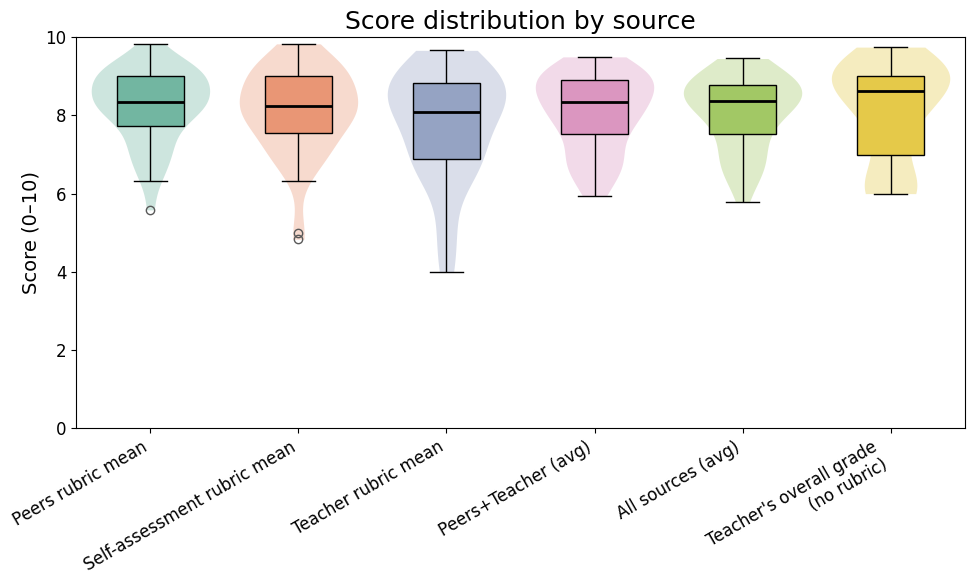}
    \caption{Distribution of scores across evaluation sources}
    \label{fig:bias}
\end{figure}

The teacher assigned significantly lower rubric scores than peers (mean difference $-0.49$, $p=0.008$) and than her own overall grade (mean difference $-0.37$, $p<0.001$). Self-assessments were significantly higher than the teacher’s rubric scores (mean difference $+0.36$, $p=0.037$), but did not differ significantly from peer scores or from the teacher's overall grade. No other comparisons yielded significant differences. Figure~\ref{fig:bias} illustrates the distribution of grades across evaluation sources, highlighting the tendency for peer and self-assessments to produce slightly higher scores than the teacher’s rubric-based evaluations.

\begin{table*}[h]
\centering
\caption{Paired-sample tests of mean differences between evaluators (0--10 scale, $n=46$). 
\textit{Note.} * $p<.05$, ** $p<.01$, *** $p<.001$.}
\label{tab:bias}
\begin{tabular}{p{6.5cm}ccccc}
\hline
\textbf{Comparison} & \textbf{Mean A} & \textbf{Mean B} & \textbf{Diff (A--B)} & \textbf{Test} & \textbf{$p$} \\
\hline
Mean rubric (Teacher) vs. Mean rubric (Peers)        & 7.76 & 8.25 & $-0.49$ & Wilcoxon, $W=267.5$ & 0.008** \\
Mean rubric (Teacher) vs. Teacher’s overall grade    & 7.76 & 8.13 & $-0.37$ & $t(45)=-3.73$ & $<0.001$*** \\
Mean rubric (Peers) vs. Teacher’s overall grade      & 8.25 & 8.13 & $+0.12$ & $t(45)=0.89$  & 0.38 \\
Mean rubric (Teacher + Peers) vs. Teacher’s overall grade & 8.09 & 8.13 & $-0.04$ & $t(45)=-0.45$ & 0.66 \\
Self-assessment vs. Mean rubric (Peers)              & 8.12 & 8.25 & $-0.14$ & $t(45)=-0.93$ & 0.36 \\
Self-assessment vs. Mean rubric (Teacher)            & 8.12 & 7.76 & $+0.36$ & $t(45)=2.15$  & 0.037* \\
Self-assessment vs. Teacher’s overall grade          & 8.12 & 8.13 & $-0.01$ & $t(45)=-0.09$ & 0.93 \\
Mean rubric (Teacher + Peers + Self) vs. Teacher’s overall grade & 8.09 & 8.13 & $-0.04$ & $t(45)=-0.36$ & 0.72 \\
\hline
\end{tabular}
\end{table*}

\section{Discussion}\label{s:discussion}
\subsection{Agreement Across Evaluators}
The results indicate that when comparing evaluators (Table~\ref{tab:agreement}), the level of agreement between evaluators was substantial according to Gwet’s AC2 coefficients and relatively consistent across all comparisons, with slightly higher values observed between peer evaluators and between peers and self-assessments. In contrast, the lowest agreement was found between the teacher and the presenters’ self-assessments, a pattern commonly reported in peer assessment literature where students tend to evaluate themselves more leniently than teachers do \cite{tang2023comparison,ross2006reliability,kilic2016examination}.

It is important to note that weighted Cohen’s $\kappa$ values were considerably lower, which can be explained by the well-known \textit{kappa paradox}, a phenomenon where high observed agreement coexists with low $\kappa$ values due to prevalence or bias effects \cite{zec2017high}. For this reason, Gwet’s AC2 provides a more robust and interpretable measure of inter-rater agreement in this context.

Regarding overall concordance among all evaluators (Table~\ref{tab:kendall}), both Kendall’s W and global AC2 values indicate substantial agreement across groups, suggesting that while individual raters may differ in their strictness, their relative ranking of performance was largely consistent.

An important factor that may have contributed to this level of consistency is the use of a detailed rubric within the AICoFe system, which included clearly defined levels of scale. The rubric was accessible to all participants throughout the course and explicitly discussed during training sessions, fostering a shared understanding of the assessment criteria. This common framework likely helped align interpretations across teacher, peers, and self-assessors, reducing subjectivity and enhancing reliability. Moreover, the substantial agreement observed between peers and between peers and the teacher suggests that students were able to apply the evaluation criteria in a manner broadly consistent with the teacher’s judgments. These findings support previous evidence that well-structured rubrics and transparent assessment procedures can enable peer evaluations to approximate teacher assessments in reliability \cite{kilic2016examination,tang2023comparison}.

\subsection{Correlation Across Evaluations}
The correlation analyses offer further evidence of the validity and coherence of the evaluation process. As shown in Table~\ref{tab:correlation}, the teacher’s rubric-based scores correlated very strongly with her overall grade ($r = 0.88$), confirming internal consistency in the teacher’s judgment. Peer assessments were also significantly correlated with the teacher’s overall grade ($r = 0.69$), indicating that students were generally able to recognize similar performance patterns. This value is comparable to the average correlation reported in large-scale meta-analyses of peer and teacher assessments, which range from moderate to substantial across diverse educational contexts \cite{falchikov2000student,li2016peer}.

In contrast, self-assessments showed a moderate correlation with the teacher’s grade ($r = 0.59$), suggesting that students’ evaluative judgments were partially aligned but still affected by calibration and self-perception biases. This pattern is consistent with previous research showing that students often misjudge their own performance levels, a phenomenon sometimes explained by the Dunning–Kruger effect, whereby lower-performing individuals tend to overestimate their abilities while higher-performing individuals underestimate them \cite{kruger1999unskilled, dunning2011dunning}. Similar trends, strong peer–self relationships but weaker alignment with teacher ratings, have also been reported in professional education contexts \cite{mehrdad2012comparative}.

When examining correlations among rubric-based mean scores (Table~\ref{tab:correlation_rubrics}), all coefficients were positive and statistically significant ($r = 0.54$–$0.62$). These values suggest that, despite small individual differences in scoring strictness, evaluators applied the rubric in a consistent way to rank performance.

\begin{figure*}[th]
    \centering
    \includegraphics[width=\linewidth]{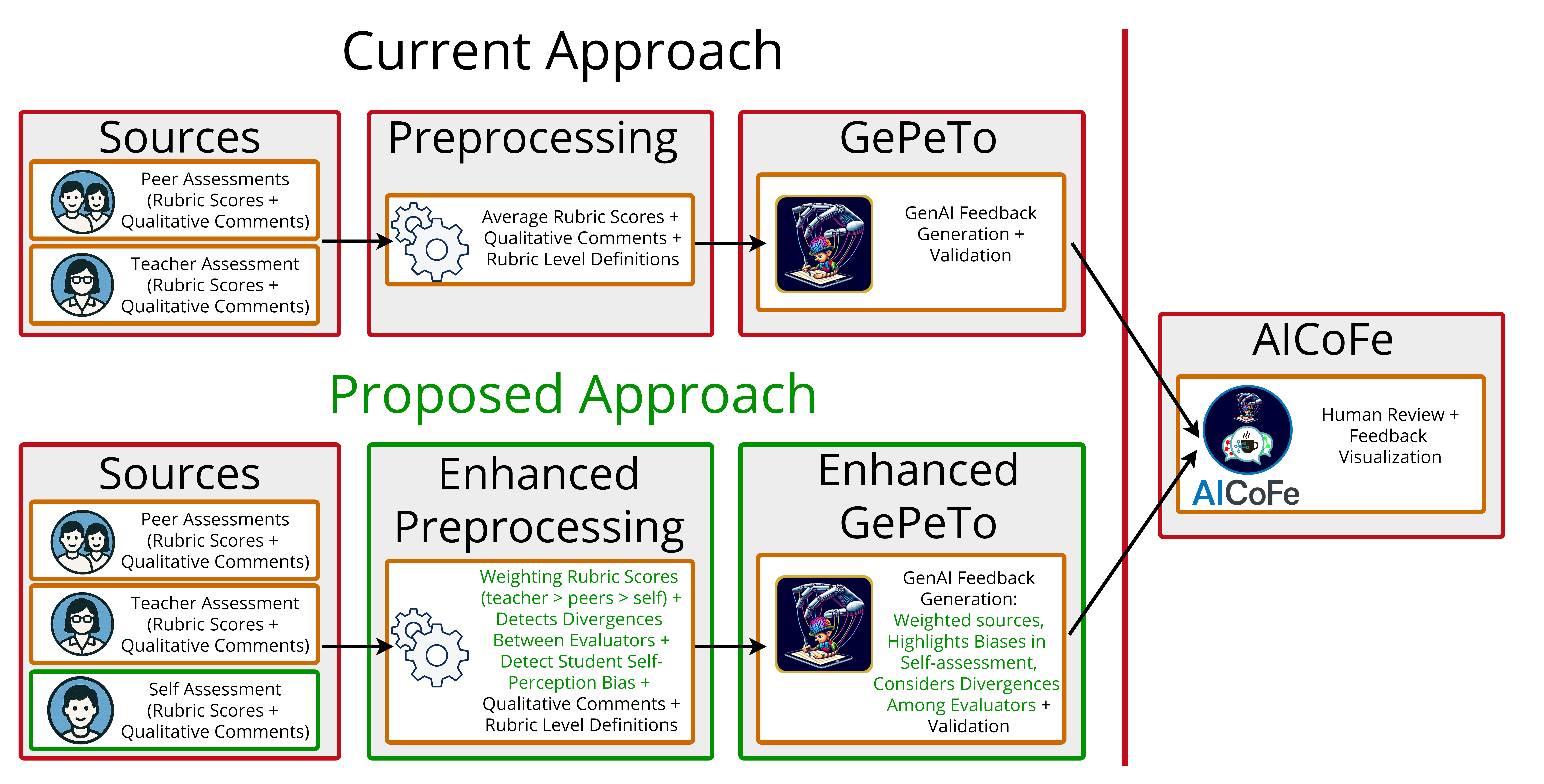}
    \caption{Comparison between the Current Approach (top) and the Proposed Enhanced Approach (bottom) for Generative AI–based feedback within the AICoFe system}
    \label{fig:proposed_approach}
\end{figure*}

\subsection{Bias Analysis}

The comparative analysis revealed that students tended to assign slightly higher scores to themselves and their peers than those assigned by the teacher, a pattern that has been consistently reported in previous studies \cite{oren2018self}. However, the magnitude of this bias was limited, and differences between peer and self-assessments were not statistically significant (Table~\ref{tab:bias}), suggesting that students’ scoring tendencies were relatively homogeneous.

The higher mean scores in peer and self-assessments reflect a general positive scoring bias, where raters systematically overestimate performance \cite{falchikov2000student,patri2002influence}. In our study, the absence of strong systematic deviations between peer and self-assessment scores, combined with the moderate agreement indices (AC2 = 0.63), suggests that bias was more related to overall score inflation than reciprocal favoritism. These findings are consistent with previous evidence indicating that reciprocity or collusion effects in peer assessment are typically negligible \cite{magin2001reciprocity}.

Interestingly, the teacher’s rubric scores were consistently lower than her own overall grades, which suggests a stricter interpretation of the detailed rubric criteria but a more holistic and lenient approach when assigning the final global grade.

\subsection{Implications for Generative AI-Enhanced Feedback}

The findings of this study offer actionable insights for the evolution of generative AI–based feedback systems such as AICoFe \cite{becerra2025enhancing}. The current implementation of AICoFe integrates teacher and peer assessments as inputs for the generative model GePeTo \cite{becerra2024generative}, which produces personalized feedback grounded in rubric-based evaluations. This AI-generated feedback is presented through Learning Analytics dashboards to support students’ self-reflection and improvement. To ensure pedagogical validity and ethical oversight, all generative outputs are reviewed by teachers before being shared with students.

However, as illustrated in Figure~\ref{fig:proposed_approach}, the current configuration of GePeTo relies on averaged rubric scores and aggregated qualitative comments from teacher and peer sources. This uniform weighting assumes equivalent reliability among evaluators, overlooking systematic divergences identified in our analyses—for example, that teachers apply stricter criteria while peers tend to provide more lenient ratings. Such discrepancies highlight the need for a more nuanced integration mechanism that accounts for evaluator bias and relative expertise.

To address these limitations, we propose an enhanced approach (see Figure~\ref{fig:proposed_approach}) that incorporates the following elements:
\begin{itemize}
\item Weighted integration of evaluator inputs, assigning greater influence to teacher scores, moderate weight to peer ratings, and including self-assessment as an additional source.
\item Bias and divergence detection during preprocessing, enabling the model to identify patterns such as overestimation in self-assessments or leniency among peers.
\item Enhanced generative reasoning within GePeTo, whereby the AI feedback generation process explicitly considers divergences among evaluators and highlights potential perception biases in its explanations.
\end{itemize}

In this proposed design, GePeTo evolves from a purely data-driven aggregator into a context-aware feedback mediator capable of integrating human evaluations through pedagogically informed weighting and interpretive bias awareness.

\section{Conclusions and Future Work}\label{s:conclusions}

This study examined the relationships among teacher, peer, and self-assessments within the AICoFe system to inform the design of Generative AI–enhanced feedback. Results revealed moderate to substantial agreement among evaluators and significant correlations with the teacher’s overall grade, supporting the reliability of peer and self-assessment data when guided by a well-defined rubric. However, systematic differences were identified such as teachers applying stricter criteria and students showing a mild positive bias, revealing that each evaluative source contributes distinct perspectives and biases, underscoring the importance of modelling these differences.

Building on these insights, we proposed an enhanced configuration of the GenAI model in AICoFe that integrates weighted evaluator inputs, bias detection, and context-aware generative reasoning. This design moves beyond data averaging toward a human-informed process that reflects both performance evidence and evaluative perspective.

The analyses and enhancement proposal for the GePeTo model presented in this study aim not only to improve the quality and reliability of AI-assisted feedback but also to establish a generalizable framework for the evaluation of key competencies in Engineering Education. By extending the proposed framework and methodology to diverse skill domains, our approach may evolve into a robust and adaptable instrument for assessing professional and technical abilities, particularly those that foster critical thinking, problem-solving, and reflective judgment, thus contributing to the holistic development of engineering students.

Future work will focus on validating this enhanced model through classroom studies, examining its impact on feedback quality, student engagement, and calibration of self-assessment skills. In this regard, it will be especially relevant to investigate how different forms of self-assessment guidance interact with GenAI-enhanced feedback. Prior research has shown that analytic rubrics can increase performance but may also reduce learning-oriented self-regulation, whereas self-assessment scripts tend to foster metacognitive monitoring and process-focused reflection, even if they offer less explicit guidance on expected performance standards \cite{panadero2014rubrics}. Building on these findings, future iterations of AICoFe could experiment with alternative or hybrid scaffolds (e.g., rubric-only, script-based, or dynamically adapted combinations) and examine their differential effects on students’ self-regulation, perceived workload, and bias patterns in peer and self-assessment.

In addition, we plan to extend AICoFe’s data integration layer by incorporating multimodal learning analytics \cite{becerra2025mosaicf}, including biosensor \cite{becerra2025enhancing_review} and behavioral data, to capture affective and physiological dimensions of learning \cite{becerra2025ai}.

\section*{Acknowledgment}
Support by projects: Cátedra ENIA UAM-VERIDAS en IA Responsable (NextGenerationEU PRTR TSI-100927-2023-2), M2RAI (PID2024-160053OB-I00, MICIU/FEDER) and SNOLA (RED2022-134284-T). Alvaro Becerra is funded by a predoctoral contract (FPI) from the Comunidad de Madrid (PIPF-2024/COM-34288).

\bibliographystyle{unsrt}
\bibliography{bibliography} 

\end{document}